**GPTON – Generative Pre-trained Transformers enhanced with Ontology Narration for accurate annotation of biological data**

Rongbin Li[1,*], Wenbo Chen[1,*], Jinbo Li[1], Hanwen Xing[1], Hua Xu[2], Zhao Li[1,&], W. Jim Zheng[1,&]

[1] McWilliams School of Biomedical Informatics, University of Texas Health Science Center at Houston, Houston, TX 77030
[2] Section of Biomedical Informatics and Data Science, School of Medicine, Yale University, New Haven, CT 06510, United States

* These authors contributed equally

& Corresponding authors: Zhao.Li@uth.tmc.edu, Wenjin.J.Zheng@uth.tmc.edu

**Abstract**

By leveraging GPT-4 for ontology narration, we developed GPTON to infuse structured knowledge into LLMs through verbalized ontology terms, achieving accurate text and ontology annotations for over 68% of gene sets in the top five predictions. Manual evaluations confirm GPTON's robustness, highlighting its potential to harness LLMs and structured knowledge to significantly advance biomedical research beyond gene set annotation.

**Main**

Generative Pre-trained Large Language Models (LLMs), trained on vast datasets (e.g., over 15 trillion tokens for Llama 3 models[1]), can process and infer complex information relationships beyond traditional methods. This ability is particularly useful in biology for tasks like annotating gene sets, overcoming conventional analysis limits[2]. However, current applications of large language models (LLMs) in gene set analysis often fall short because they typically generate ontology labels directly, resulting in suboptimal outcomes.[3-5] This poor performance stems from the fact that ontologies are neither widely nor precisely used in the biomedical literature—a major component of the training data for LLMs.[4, 5] Consequently, while LLMs excel at generating natural language, they struggle with structured, specialized data like ontologies due to the lack of context and understanding in their training data.

To address this challenge, we developed GPTON, Generative Pre-trained Transformers enhanced with Ontology Narration, for accurate annotation of biological data. Ontology narration, or ontology verbalization, converts ontology terms into natural language without losing their meaning[6]. Fine-tuning LLMs with verbalized ontology terms has the potential to enhance their performance in accurately annotating gene sets—a pivotal task in analyzing single-cell RNA sequencing and other high-throughput 'omics' experiments. GPTON overcomes limitations of traditional gene set analysis methods, such as Gene Set Enrichment Analysis (GSEA), which rely on existing gene set databases and single-gene annotations that are often incomplete and limited in scope[7]. Additionally, these traditional methods cannot generate summary text to describe gene set functions, hindering the discovery of novel gene interactions and pathways.

GPTON uses an innovative three-step method for gene set annotation with LLMs: 1) Ontology Narration: Utilizing GPT-4 to verbalize GO terms from the Biological Process branch into narratives (**Fig 1a**); 2) LLM Fine-tuning: Using these narratives to label gene sets from MSigDB and fine-tune LLMs like Llama 3 for generating aligned narrative summaries (**Fig 1b**); and 3) Mapping Narrative Summaries: Mapping narrative summaries to verbalized GO terms using cosine similarity with an encoding module[8] (**Methods**), offering GO term annotations for each gene set (**Fig 1c**). We extensively evaluated GPTON using the Llama 3 8B and 70B models for annotating human and mouse gene sets (**Methods**). For comparison, we included GeneAgent[2] and GPT-4[3], which are current state-of-the-art LLM-based approaches for identifying GO terms for gene set annotation (**Methods**). Hyperparameter optimization was performed using grid search to determine the optimal learning rate, LoRA size, and batch size. We assessed performance using widely recognized metrics, such as ROUGE scores[9] for word-level similarity and BERTScore[10] for semantic similarity between generated gene set labels and ground truth labels (**Methods**). Additionally, domain experts conducted manual evaluation to ensure the accuracy of predicted GO terms and verbalized narratives.

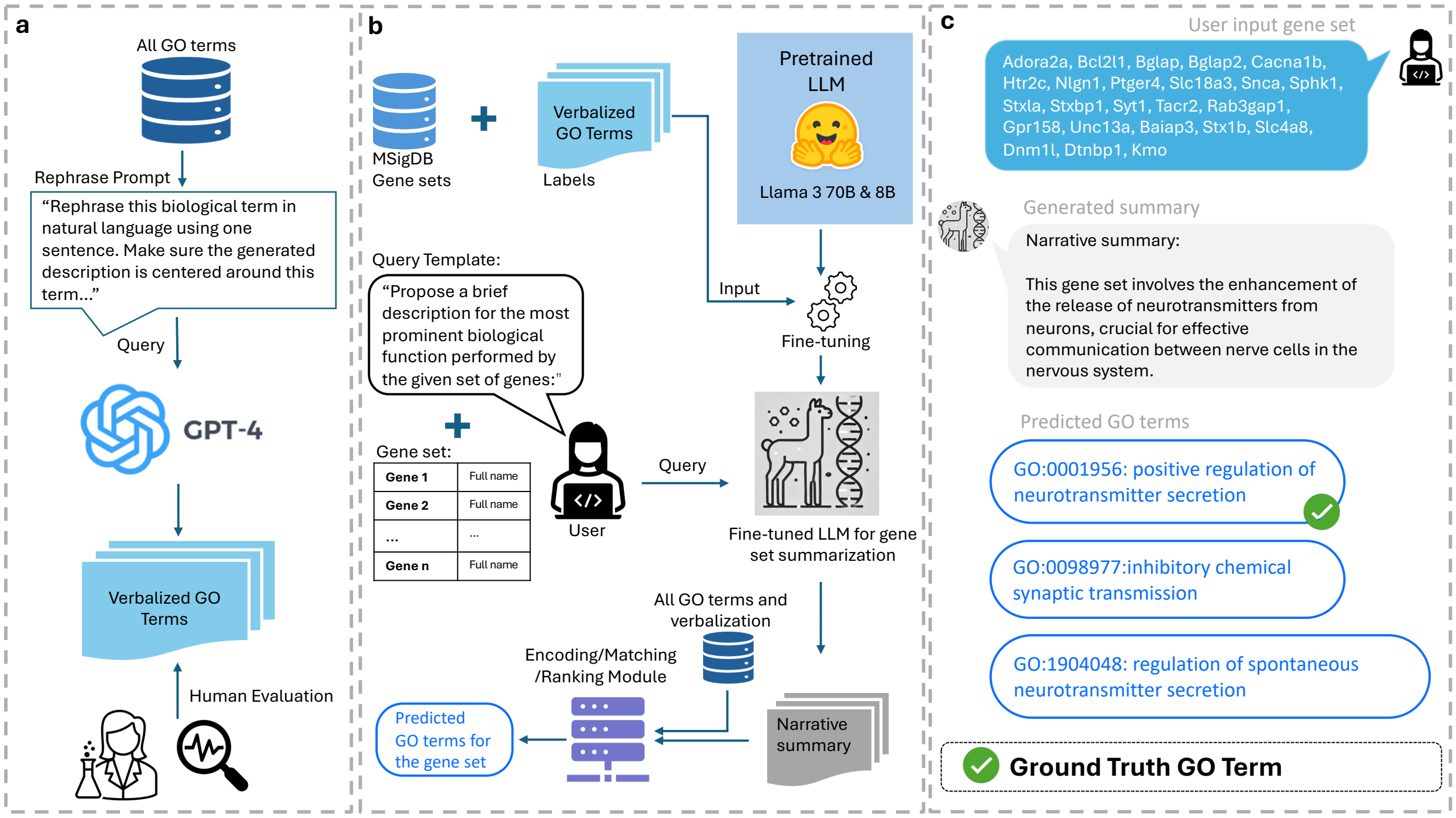

**Fig 1**. **GPTON for gene set annotation using GPT-4 verbalized ontology.** a) Workflow for GO term verbalization. b) Framework for fine-tuning LLMs with gene sets labeled with verbalized ontology, generating summaries, and mapping these summaries back to verbalized GO terms for annotation. c) Example of narrative summaries and predicted GO terms for a given gene set.

GPTON achieved 14% and 20% improvements in ROUGE-1 compared to GeneAgent and GPT-4, respectively, for human gene sets (**Fig 2a**). The GPTON model based on Llama 3 70B demonstrated the best performance across all four metrics for both human and mouse gene sets (**Fig 2a, 2b**, **Supplementary Table 1, 2** and **Supplementary Fig 1**). While directly fine-tuning Llama 3 with GO labels outperformed the base model, it remained suboptimal compared to models fine-tuned with verbalized ontology (**Fig 2b, Supplementary Table 2** and **3**).

We selected GPTON based on the Llama 3 8B model for further testing and analysis due to its relatively compact size and strong performance. The stability of GPTON is demonstrated by the consistent ROUGE-1 scores across 5-fold cross-validation (**Fig 2c** and **Supplementary Table 4**). This stability remains true across all four metrics **(Supplementary Fig 2** and **Supplementary Table 4)**.

GPTON demonstrated high consistency and accuracy. Two bioinformaticians manually evaluated the agreement between the original GO terms assigned to each gene set (A) and those predicted by GPTON (B) (**Methods**). For the top 50 gene sets, 92% of the predicted GO terms exactly matched or were highly relevant to the original ones, compared to 69% for the bottom 50 gene sets (**Fig 2d**). The alignment between GO terms and their GPT-4 verbalized narratives (C) was also evaluated, with 98% of the verbalized narratives for the top 50 gene sets and 92% for the bottom 50 sets accurately aligning with the original GO terms in meaning. The assessment was objective, based on whether a GO term for a gene set could be mapped back to its ground truth annotation.

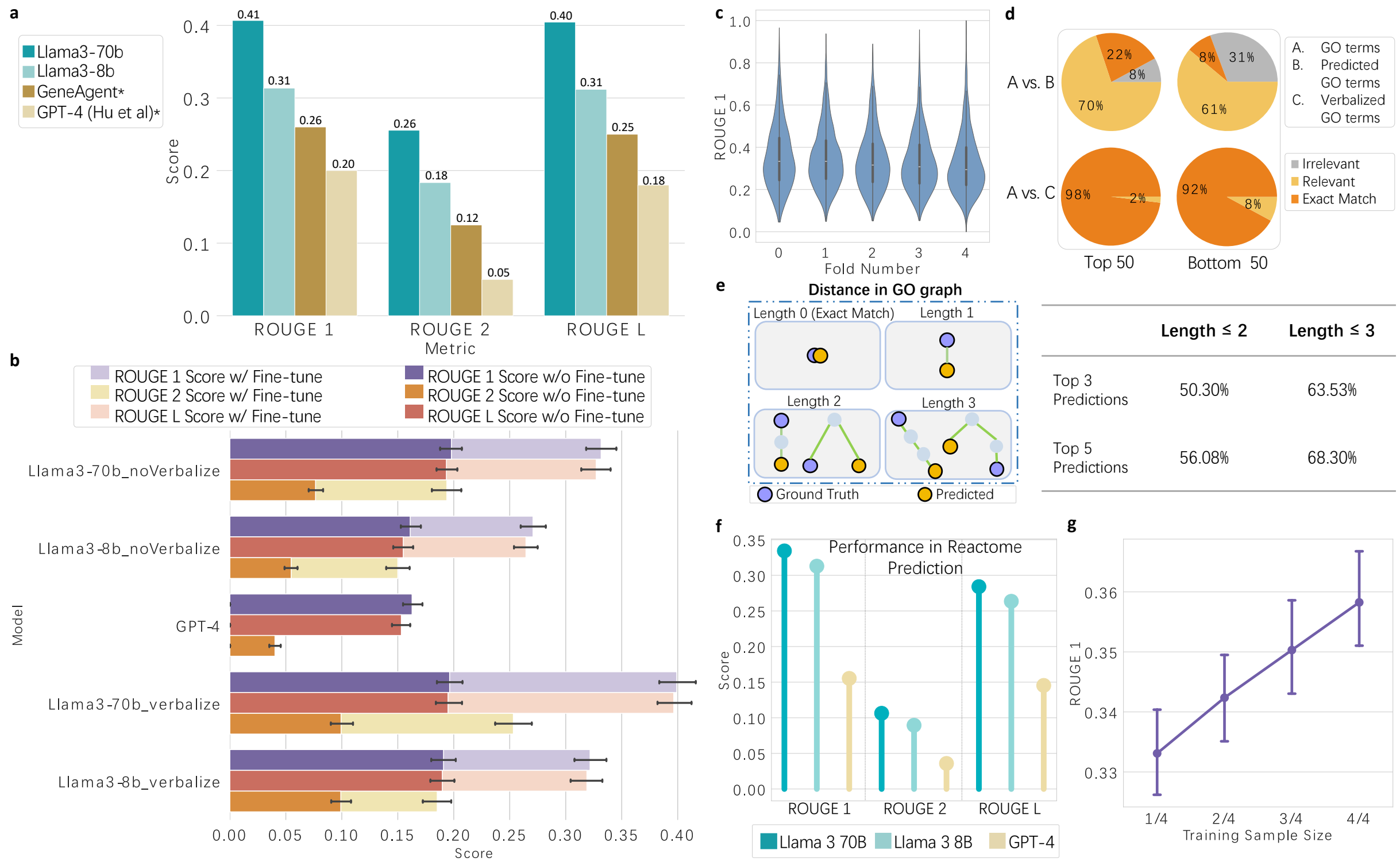

**Fig 2**. **GPTON significantly improves gene set annotation.** a) ROUGE scores of GPTON, including Llama 3 70B and 8B fine-tuned with verbalized GO terms, compared to GeneAgent and GPT-4 for human gene sets annotation. b) ROUGE scores of Llama 3 models fine-tuned with verbalized ontology terms (verbalize) versus original ontology terms (noVerbalize), along with GPT-4 model for direct GO term generation for mouse gene sets annotation. The error bars represent 95% confidence intervals. c) Consistent performance of the fine-tuned Llama 3 8B model evaluated through 5-fold cross-validation on mouse gene sets. The thick lines represent interquartile ranges, and thin lines are 1.5x interquartile ranges. d) Human evaluation of GPTON's correct and relevant GO term predictions (A vs. B row) and consistent GO term verbalization (A vs. C row) in 100 selected gene sets. e) Accuracy of GPTON predictions among top-3 and top-5 predicted GO terms as evaluated by GO graph topology. f) Performance of fine-tuned Llama 3 70B and 8B models on the mouse Reactome dataset. g) Performance of the Llama 3 8B model with different training sample sizes. The error bars represent 95% confidence intervals.

We further evaluated GPTON for gene set annotation using the topology of the GO graph. Our results show that GPTON can identify exact or highly relevant GO terms in over 68% of mouse gene sets among the top five predictions (**Fig 2e**). A highly relevant GO term is defined as being within three or fewer edges from the ground truth GO term in the GO graph (**Fig 2e**). This quantitative evaluation, free from potential human subjectivity, further underscores GPTON's utility in real-world gene set analysis tasks, demonstrating its ability to generate highly informative results suitable for manual confirmation.

To examine GPTON's generalizability, we predicted annotations for external sets of 1,670 human and 1,247 mouse gene sets from Reactome (**Methods**). GPTON, based on two fine-tuned Llama 3 models, achieved ROUGE-1 scores of 0.33 and 0.31 on the mouse dataset, respectively, significantly surpassing the baseline GPT-4 model's score of 0.16 (**Fig 2f** and **Supplementary Table 5**). On the human dataset, the improvement reaches 19% compared to GPT-4 (**Supplementary Fig 3** and **Supplementary Table 6**). These results demonstrate GPTON's strong generalizability across datasets of different origins, even those not seen during training. This is crucial for annotating gene sets with few or no existing ontology labels and for providing accurate annotations for novel gene sets from experimental investigations. Furthermore, this capability can be further improved by additional training samples. When the training set was at ¼, ½, and ¾ of its original size, there is a consistent increase in performance in generating narrative summaries with larger training sets (**Fig 2g, Supplementary Fig 4** and **Supplementary Table 7)**.

GPTON pioneers a flexible and informative approach to integrating structured knowledge into LLMs, delivering reliable annotations for gene sets even without existing ontology labels or from novel experiments. Mapping generated summaries back to structured knowledge minimizes hallucination, a common LLM limitation. Expanding training data with additional resources and structured knowledge could enhance GPTON's performance, extending its applications beyond gene set annotation to human disease research and drug discovery.

## Methodology

### *Data Collection*

We selected two species, human (Homo Sapiens) and mouse (Mus Musculus), and obtained a list of gene sets with their corresponding GO terms from MSigDB[11] database for both species to conduct our experiments. We also conducted filtering on the two lists of gene sets with the following criteria:

- The collection category has to be a GO Biological Process (GO:BP);
- The gene sets must have an identified exact source;
- The number of genes contained in each gene set must be less than 500.

The filtered dataset contains 7,423 and 7,292 gene sets for mouse and human, respectively. For the collected gene sets, we also obtained the description of each gene from the NCBI gene info table[12]. For the human dataset, we used the same evaluating gene sets in GeneAgent[5] as the test set. To prevent potential data leakage, we filtered out gene sets in the human training dataset that had the same GO term labels or genes as those in the test set. The mouse dataset was separated into training and test sets by an 8:2 ratio. Our study then revolved around fine-tuning LLMs with the collected gene sets and their labels, such that the LLMs are deployed to the task of generating the gene set annotations.

A second dataset was also constructed from the Reactome subset of the MSigDB database with the collection category being: CP:REACTOME[11], which contains 1,670 human gene sets and 1,247 mouse gene sets after we use a similar filtering criteria to obtain labels of the Reactome gene sets.

*Gene Ontology Verbalization*

Prior to the LLM fine-tuning, we first converted the collected GO terms from short phrases to natural language sentences, extending their length and context. This was done by querying GPT-4 (GPT4-0125) to generate verbalized narratives for all GO terms. During this process, we only changed the maximum number of tokens to 1,024 and kept all other parameters as default for GPT-4. **Box M1** shows the query template we used:

**Box M1. Prompt for ontology verbalization**

**System prompt:**
You are a biologist who helps people find information.

**User prompt:**
Rephrase this biological term in natural language using one sentence. Make sure the generated description is centered around this term and within 30-50 words.

The rephrased sentence should first identify which category, out of biological process, cellular component, and molecular function, the given term is from, and then provide an explanation of this term.

Be concise, do not use unnecessary words.

Don't include gene symbol information in your rephrased description.

Here are several examples:

Biological Term: pole plasm
Rephrased Description: This term refers to a cellular component which is a specialized area in the egg cytoplasm that contains critical materials needed for the early stages of embryonic development and the formation of germ cells.

Biological Term: mitochondrial fragmentation involved in apoptotic process
Rephrased Description: This term refers to a biological process which is the splitting of mitochondria into smaller parts, which is essential for the programmed death of a cell.

Biological Term: guanylate cyclase regulator activity
Rephrased Description: This term refers to a molecular function which is the control of the enzyme guanylate cyclase, which is vital for converting GTP to cGMP, a key signaling molecule in various cellular processes.

The biological term is: [GO term].

*LLM Selection and Prompting Pipeline*

To evaluate the ability of generative LLMs in predicting gene set annotation labels after fine-tuning and find the model that is most efficient and effective, we selected two different versions of the state-of-the-art LLMs hosted on Hugging Face: Llama 3 8B, and Llama 3 70B[1].

We then composed a prompt template to fine-tune the Llama 3 models while using the verbalized GO terms generated by GPT-4 as ground truth. **Box M2** shows the query template.

*Model Training*

**Box M2. Prompt for fine-tuning LLM:**

**System prompt:**
You are a senior biologist.

**User prompt:**
Propose a brief description for the most prominent biological function performed by the given set of genes. Be concise, do not use unnecessary words. Be specific, avoid overly general statements such as "the genes are involved in various cellular processes". Be factual, do not editorialize.

The genes and corresponding descriptions are: [Genes and descriptions]

Training data containing the proposed prompts as input and verbalized GO terms as ground truth labels were generated and utilized to fine-tune the LLMs. During model training, 1/8 of the training samples were further separated as the validation set, while the remaining 7/8 of data were learned upon.

The Transformer Reinforcement Learning (TRL) library from Hugging Face[13] was used to accommodate our usage of the NVIDIA H100 GPUs. Since full fine-tuning on larger models such as Llama 3 70B is not feasible with our hardware resources, we employed two fine-tuning strategies: full-model fine-tuning (Llama 3 8B) and LoRA with Quantization (QLoRA) (Llama 3 8B and 70B)[14]. To draw out the most optimal performance from LLMs, we also conducted hyperparameter searching using Llama 3 8B model by designating learning rate, LoRA parameters (rank and alpha values), and batch size to tweak, and the values we decided upon are 0.00001, [128,512], 16, respectively. The Llama 3 70B model adopted the same hyperparameters as those optimized for the 8B model.

*Identify GO terms using generated summary by embedding similarity*

The target GO term for each gene set was identified by comparing the semantic embeddings of the verbalized narrative for all GO term candidates with the narrative summary generated by the fine-tuned LLM for the given gene set. We adapted MedCPT[8], a Transformer-based encoder model pre-trained on PubMed query-article pairs for zero-shot semantic information retrieval in biomedicine, to embed both the generated gene set description and the verbalized narratives of all GO terms in the biological process namespace. The cosine similarity between the gene set summary and each candidate GO term was then calculated as the dot product of the normalized vectors. The GO terms corresponding to the highest-ranked verbalized narratives were retrieved as the predicted target GO terms for the gene set.

*Benchmarking Methods*

GeneAgent

GeneAgent[2] is an LLM-based approach designed to enhance gene set analysis by minimizing the common issue of hallucinations in LLMs. Built upon GPT-4, GeneAgent operates through a structured workflow of generation, self-verification, modification, and summarization. Initially, it generates process names and analytical texts for input gene sets. It then employs a selfVeri-Agent to autonomously verify these outputs against curated knowledge from domain-specific databases through Web APIs. This verification involves extracting claims and comparing them with database information, iteratively refining the process names and analytical narratives to ensure accuracy. GeneAgent's self-verification capability allows it to discern and correct hallucinations, improving reliability. Comparative studies reveal that GeneAgent outperforms standard GPT-4 in generating accurate biological process names and provides more informative gene set annotations.

GPT-4

Hu et al.[3] evaluated the ability of various LLMs to uncover the functions of gene sets derived from functional genomics experiments. Their method involves querying a panel of LLMs, including GPT-4, Gemini-Pro, Mixtral-Instruct, and Llama 2 70B, to directly generate GO term names, supporting analysis texts, and confidence scores for the given gene set. The process includes benchmarking LLMs against gene sets from the Gene Ontology to assess their accuracy and exploring their ability to propose novel functions for gene sets derived from 'omics data. Notably, GPT-4 demonstrated high accuracy in recovering curated gene set names and identifying novel gene functions not reported by traditional functional enrichment methods. Therefore, the GPT-4-based approach described in this paper will serve as one of the baselines for comparative analysis in our study.

Specifically, we used the following prompt to query the GPT-4 model to directly predict GO terms. This prompt template includes a one-shot example shown in **Box M3**. The scores of GeneAgent and GPT-4 are estimated based on the graph reported in GeneAgent paper since they didn't report their numerical results.

**Box M3. Prompt for one-shot query of GPT-4:**

**System prompt:**
You are a biologist who helps people find information.

**User prompt:**
Propose a brief description for the most prominent biological function performed by the given set of genes. Be concise, do not use unnecessary words.

Be specific, avoid overly general statements such as "the genes are involved in various cellular processes". Be factual, do not editorialize.

Here is one example of the input and response:

"Input": C3: complement component 3; C6: complement component 6; Cd59a: CD59a antigen; Cfh: complement component factor h; Cd59b: CD59b antigen

"Response": activation of membrane attack complex

The genes and corresponding descriptions are: [Gene descriptions]

*Evaluation Metrics*

The performance of the LLMs was evaluated by comparing the generated response with the ground truth in the testing set and calculating four widely used metrics: ROUGE-1 score, ROUGE-2 score, ROUGE-L score[9], and BERTscore[10]. We utilized the Python implementation "ROUGE" (version 1.0.1) to calculate the ROUGE-1, ROUGE-2, and ROUGE-L scores. Similarly, the Python implementation "bert_score" (version 0.3.13) was imported to obtain the BERTScore.

ROUGE 1 and 2 scores represent the $1$-gram and $2$-gram recall value, respectively, between the two texts at comparison[8]. For each gene set, it is calculated by the following equation (1):

$$ROUGE\ n = \frac{\sum_{S \in references} \sum_{gram_n \in S} Count_{match}(gram_n)}{\sum_{S \in references} \sum_{gram_n \in S} Count(gram_n)} \quad (1)$$

Where $S$ is a reference summary, $gram_n$ is a $n$-gram in $S$, $Count_{match}(gram_n)$ is the number of $n$-grams that are found both in the reference summary and the generated summary, and $Count(gram_n)$ is the count of all n-grams in the reference summary[2]. In our case, $n$ can be 1 or 2.

ROUGE L score is an F1-based metric calculated from the longest common subsequence (LCS) between the two summaries. Let $X$ and $Y$ represent the reference summary and the generated summary, and $m$ and $n$ are the length of $X$ and $Y$, respectively. To obtain ROUGE L score, the precision ($P$) and recall ($R$) with respect to LCS are first calculated by the following equations (2) and (3):

$$R_{LCS} = \frac{LCS(X,Y)}{m} \quad (2)$$

$$P_{LCS} = \frac{LCS(X,Y)}{n} \quad (3)$$

Where $LCS(X, Y)$ is the length of the longest common subsequence between $X$ and $Y$. ROUGE L score is then calculated by:

$$ROUGE\ L = \frac{(1+\beta^2) R_{LCS} \cdot P_{LCS})}{R_{LCS} + \beta^2 \cdot P_{LCS}} \quad (4)$$

Where $\beta$ is a constant hyperparameter value.

BERTScore is also a F1-based metric, where the $P$ and $R$ are calculated by the matrix similarity between the tokens in two summaries[10]. Let $x$ and $y$ be the tokens in $X$ and $Y$, where $X$ and $Y$ are defined as above, the following equations calculate the $R$, $P$ and BERTScore:

$$R_{BERT} = \frac{1}{|x|} \sum_{x_i \in x} max_{y_j \in y} (x_i^{\top} y_j) \quad (5)$$

$$P_{BERT} = \frac{1}{|y|} \sum_{y_j \in y} max_{x_i \in x} (x_i^{\top} y_j) \quad (6)$$

$$BERTScore = 2\frac{P_{BERT} \cdot R_{BERT}}{P_{BERT} + R_{BERT}} \quad (7)$$

*Reduced Sample Size Training*

To observe the effect that the size of the training set has on the model's performance, we tried reducing the training set by sampling multiples of quarters before fine-tuning the model and calculating its performance in generating narrative summaries in the separated test set. As the sample size partition increased from ¼ to 1, the mean accuracy of the Llama 3 70B and 8B predictions increased as well (**Fig 2g** and **Supplementary Table 9**).

*Human Evaluation*

To further assess the quality of the responses gained from LLMs in our method, we sampled 10 predictions in every 0.1 interval in terms of the ROUGE 1 score made by the Llama 3 70B model on the mouse test set. Specifically, we extracted the original GO term label (A), final predicted GO term (B), and verbalized narratives (C). We then deployed two bioinformatics experts to evaluate the semantic relevance between both the true GO term-predicted GO term pair (A vs. B consistency), and the GO term-verbalized narratives pair (A vs. C consistency). We designated three levels of consistency: irrelevant, relevant, and exact match. For a pair to be considered relevant, they must describe the same biological process, while they may slightly differ in scales and details. For example, in an A vs. C pair where the ground truth GO term is "regulation of mast cell chemotaxis", a verbalized narrative such as "This process involves the orchestration of mast cell movement…" would be considered an exact match with the GO term, while a summary such as "This process involves the enhancement of mast cell movement…" would be labeled as relevant, since although the original GO term does not specify positive or negative regulation, they are describing the same biological process at different levels of detail. Similarly, in the case of the A vs. B pair, if the reference GO term (A) is "modification of synaptic structure", then a GO term such as "regulation of synaptic structure" would be considered an exact match, while "regulation of dendritic spine morphogenesis" would be considered relevant, since dendritic spine morphogenesis is only a subset of synaptic structure.

## Data Availability

All data used in this manuscript were downloaded from publicly available sources. The gene sets and their annotation information for human gene sets was obtained from the MSigDB website (https://www.gsea-msigdb.org/gsea/msigdb/human/genesets.jsp?collection=GO:BP). The corresponding annotations for mouse gene sets are also available on the MSigDB website (https://www.gsea-msigdb.org/gsea/msigdb/mouse/genesets.jsp?collection=GO:BP). Detailed information for each gene (gene_info) was sourced from the NCBI Gene FTP database (https://ftp.ncbi.nih.gov/gene/DATA/). The basic version of the Gene Ontology OBO file which includes all Gene Ontology terms was downloaded from the Gene Ontology website (https://geneontology.org/docs/download-ontology/). The human Reactome dataset was downloaded from the MSigDB database under the collection category CP (https://www.gsea-msigdb.org/gsea/msigdb/human/genesets.jsp?collection=CP:REACTOME). Similarly, the mouse Reactome dataset was downloaded from (https://www.gsea-

msigdb.org/gsea/msigdb/mouse/genesets.jsp?collection=CP:REACTOME). We downloaded all Reactome terms from the Reactome database (https://reactome.org/download-data).

## Funding Acknowledgement

Rongbin Li is supported by NLM Training Program in Biomedical Informatics & Data Science for Predoctoral & Postdoctoral Fellows under the Gulf Coast Consortia grant 5T15LM007093-31. Jinbo Li is supported by the Cancer Prevention and Research Institute of Texas training grant RP210045. This work is also supported by the National Institutes of Health (NIH) through grants 1UL1TR003167, 1UM1TR004906-01, 1R01AG066749, 1U24MH130988-01, 5R56AG069880-02, Department of Defense W81XWH-22-1-0164, and the Cancer Prevention and Research Institute of Texas through grant RP170668 (WJZ).

## Supplementary Information

**Supplementary Table 1. Model comparison with state-of-the-art methods.** Detailed results for Fig 2a that report the mean scores and 95% confidence intervals of the models' performances, where our GPTON based on Llama 3 70B achieved the best mean score in all three metrics. The test sample set (N=990) is the same as used in GeneAgent's work[1].

| Model | Metric | Mean | 95% Confidence interval |
|---|---|---|---|
| Llama 3 70B | ROUGE 1 | **0.4067** | [0.3874, 0.4259] |
| Llama 3 70B | ROUGE 2 | **0.2556** | [0.2365, 0.2746] |
| Llama 3 70B | ROUGE L | **0.4045** | [0.3853, 0.4238] |
| Llama 3 8B | ROUGE 1 | 0.3136 | [0.2853, 0.3320] |
| Llama 3 8B | ROUGE 2 | 0.1834 | [0.1669, 0.1998] |
| Llama 3 8B | ROUGE L | 0.3119 | [0.2936, 0.3302] |
| GeneAgent* | ROUGE 1 | 0.26 | N/A |
| GeneAgent* | ROUGE 2 | 0.12 | N/A |
| GeneAgent* | ROUGE L | 0.25 | N/A |
| GPT-4 (Hu et al.)* | ROUGE 1 | 0.20 | N/A |
| GPT-4 (Hu et al.)* | ROUGE 2 | 0.05 | N/A |
| GPT-4 (Hu et al.)* | ROUGE L | 0.18 | N/A |

*The scores for GeneAgent[1] and Hu's method[2] reported here do not have confidence intervals, since the mean values are estimated from the graph reported in the GeneAgent paper because they did not provide exact numbers.

**Supplementary Table 2. Detailed model performance comparison on mouse data set for Fig 2b.** We assessed the mean performances and 95% confidence intervals of all models using the MSigDB[3] mouse gene data set (N=1489) across three different metrics. Models incorporating verbalization (labeled with "verbalize") achieved superior mean values. Additionally, one-sided t-test p-values were recorded, comparing all models' performances to the baseline GPT-4 model.

| Model | Metric | Fine-tuning | Mean | 95% Confidence interval | One-sided t-test p-value |
|---|---|---|---|---|---|
| Llama 3 70B noVerbalize | ROUGE 1 | w/ fine-tuning | 0.3316 | [0.3187, 0.3445] | 1.4926e-93 |
| Llama 3 70B noVerbalize | ROUGE 2 | w/ fine-tuning | 0.1939 | [0.1813, 0.2064] | 8.6830e-102 |
| Llama 3 70B noVerbalize | ROUGE L | w/ fine-tuning | 0.3272 | [0.3144, 0.3401] | 1.6467e-102 |
| Llama 3 70B noVerbalize | ROUGE 1 | w/o fine-tuning | 0.1979 | [0.1882, 0.2076] | 6.7063e-08 |
| Llama 3 70B noVerbalize | ROUGE 2 | w/o fine-tuning | 0.0765 | [0.0701, 0.0829] | 3.4345e-18 |
| Llama 3 70B noVerbalize | ROUGE L | w/o fine-tuning | 0.1934 | [0.1839, 0.2029] | 2.027e-10 |
| Llama 3 8B noVerbalize | ROUGE 1 | w/ fine-tuning | 0.2645 | [0.2533, 0.2757] | 2.5610e-48 |
| Llama 3 8B noVerbalize | ROUGE 2 | w/ fine-tuning | 0.1501 | [0.1400, 0.1601] | 1.0002e-76 |
| Llama 3 8B noVerbalize | ROUGE L | w/ fine-tuning | 0.2645 | [0.2533, 0.2757] | 1.2847e-53 |
| Llama 3 8B noVerbalize | ROUGE 1 | w/o fine-tuning | 0.1611 | [0.1520, 0.1701] | 5.1219e-48 |
| Llama 3 8B noVerbalize | ROUGE 2 | w/o fine-tuning | 0.0547 | [0.0489, 0.0606] | 2.0003e-76 |
| Llama 3 8B noVerbalize | ROUGE L | w/o fine-tuning | 0.1549 | [0.1462, 0.1636] | 0.0001 |
| Llama 3 70B verbalize | ROUGE 1 | w/ fine-tuning | **0.3992** | [0.3832, 0.4152] | 1.8990e-129 |
| Llama 3 70B verbalize | ROUGE 2 | w/ fine-tuning | **0.2532** | [0.2374, 0.2691] | 1.2880e-126 |
| Llama 3 70B verbalize | ROUGE L | w/ fine-tuning | **0.3965** | [0.3805, 0.4124] | 5.0104e-140 |
| Llama 3 70B verbalize | ROUGE 1 | w/o fine-tuning | 0.1963 | [0.1843, 0.2083] | 4.8711e-06 |
| Llama 3 70B verbalize | ROUGE 2 | w/o fine-tuning | 0.0993 | [0.0898, 0.1089] | 1.5504e-26 |
| Llama 3 70B verbalize | ROUGE L | w/o fine-tuning | 0.1949 | [0.1829, 0.2068] | 9.79e-09 |
| Llama 3 8B verbalize | ROUGE 1 | w/ fine-tuning | 0.3217 | [0.3069, 0.3365] | 5.1654e-70 |
| Llama 3 8B verbalize | ROUGE 2 | w/ fine-tuning | 0.1853 | [0.1718, 0.1987] | 3.1487e-82 |
| Llama 3 8B verbalize | ROUGE L | w/ fine-tuning | 0.3189 | [0.3041, 0.3337] | 5.0427e-78 |
| Llama 3 8B verbalize | ROUGE 1 | w/o fine-tuning | 0.1908 | [0.1795, 0.2021] | 5.9810e-05 |
| Llama 3 8B verbalize | ROUGE 2 | w/o fine-tuning | 0.0991 | [0.0903, 0.1079] | 1.0242e-29 |
| Llama 3 8B verbalize | ROUGE L | w/o fine-tuning | 0.1898 | [0.1785, 0.2010] | 1.4515e-07 |
| GPT-4 | ROUGE 1 | w/o fine-tuning | 0.1628 | [0.1540, 0.1715] | N/A |
| GPT-4 | ROUGE 2 | w/o fine-tuning | 0.0403 | [0.0352, 0.0454] | N/A |
| GPT-4 | ROUGE L | w/o fine-tuning | 0.1532 | [0.1449, 0.1614] | N/A |

**Supplementary Fig 1: The BERTScore performance comparison of models on mouse data set.**

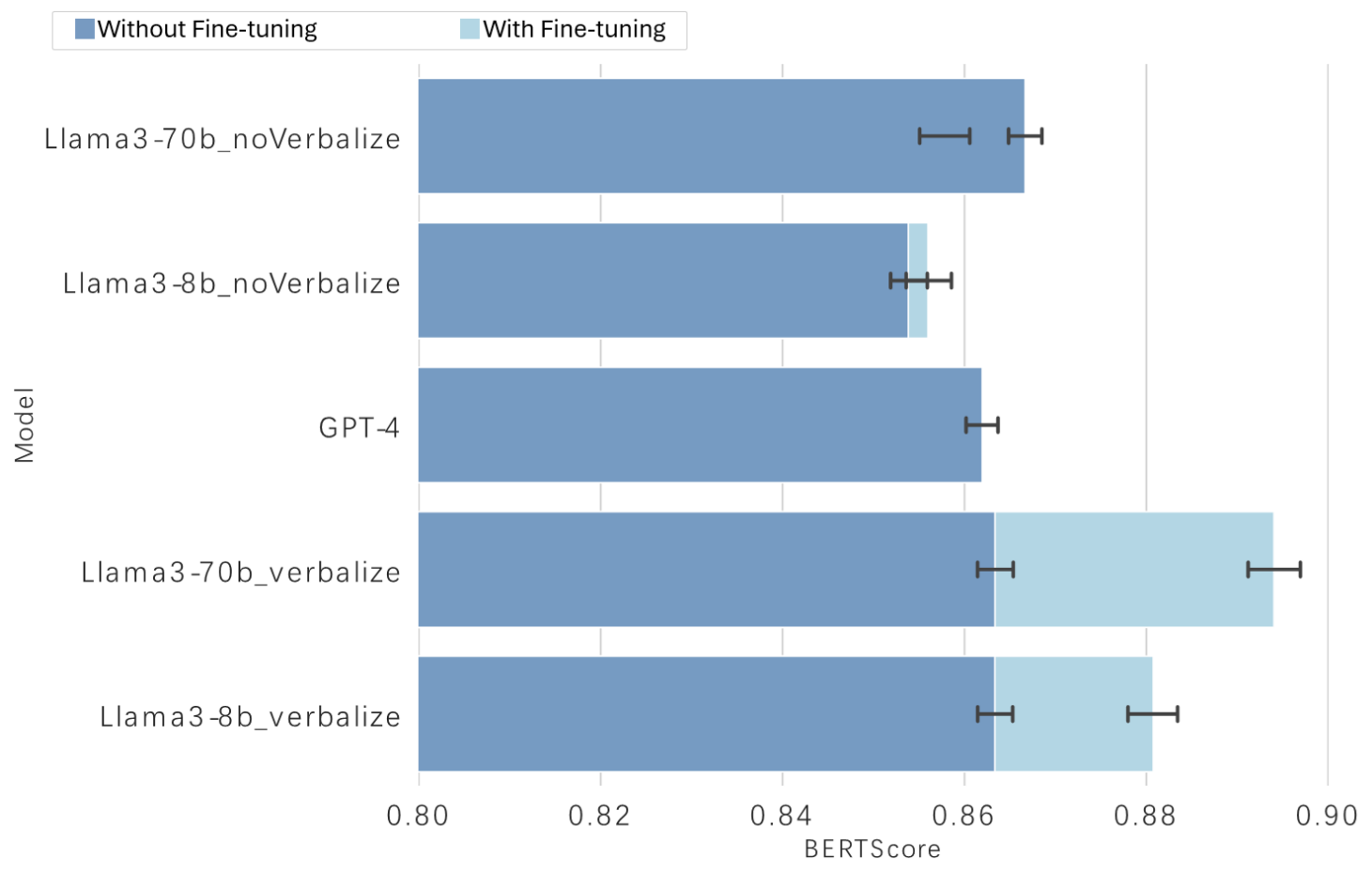

**Supplementary Fig 1. The BERTScore performance comparison of models on mouse data set.** The comparison involves models directly predicting gene set GO term annotations (noVerbalize and GPT-4) versus models predicting verbalized gene set summaries before mapping back to standard GO terms (verbalize). The models fine-tuned to directly predict GO terms perform suboptimal, while models fine-tuned to first predict verbalized GO terms provide more accurate GO term predictions at the final step. The best performing model, Llama 3 70B fine-tuned with verbalized GO terms, achieves a BERT score of 0.89. The error bars represent 95% confidence intervals. With detailed values and parameters available in **Supplementary Table 3**.

**Supplementary Table 3. Model performance comparison on mouse data set for Supplementary Fig 1**. We tested various models using a mouse gene dataset (N=1489) and reported their mean BERTScores with 95% confidence intervals. Models incorporating verbalization ("verbalize") achieved higher mean values. P-values of one-sided t-tests are also recorded, where all models' performances were compared to the baseline GPT-4 model.

| Model | Metric | Fine-tuning | Mean | 95% Confidence interval | One-sided t-test p-value |
|---|---|---|---|---|---|
| Llama 3 70B noVerbalize | BERTScore | w/ fine-tuning | 0.8577 | [0.8548, 0.8606] | 0.0638 |
| Llama 3 70B noVerbalize | BERTScore | w/o fine-tuning | 0.8667 | [0.8648, 0.8685] | 3.6707e-06 |
| Llama 3 8B noVerbalize | BERTScore | w/ fine-tuning | 0.8560 | [0.8535, 0.8585] | 0.9991 |
| Llama 3 8B noVerbalize | BERTScore | w/o fine-tuning | 0.8538 | [0.8518, 0.8559] | 1.0000 |
| Llama 3 70B verbalize | BERTScore | w/ fine-tuning | **0.8940** | [0.8913, 0.8968] | 8.5863e-85 |
| Llama 3 70B verbalize | BERTScore | w/o fine-tuning | 0.8634 | [0.8613, 0.8654] | 0.0316 |
| Llama 3 8B verbalize | BERTScore | w/ fine-tuning | 0.8808 | [0.8783, 0.8833] | 4.7336e-37 |
| Llama 3 8B verbalize | BERTScore | w/o fine-tuning | 0.8633 | [0.8614, 0.8653] | 0.0329 |
| GPT-4 | BERTScore | w/o fine-tuning | 0.8609 | [0.8591, 0.8626] | N/A |

**Supplementary Fig 2. The violin plots of the Llama 3 8B model performance in different metrics across 5-fold cross-validation using mouse and human data sets.**

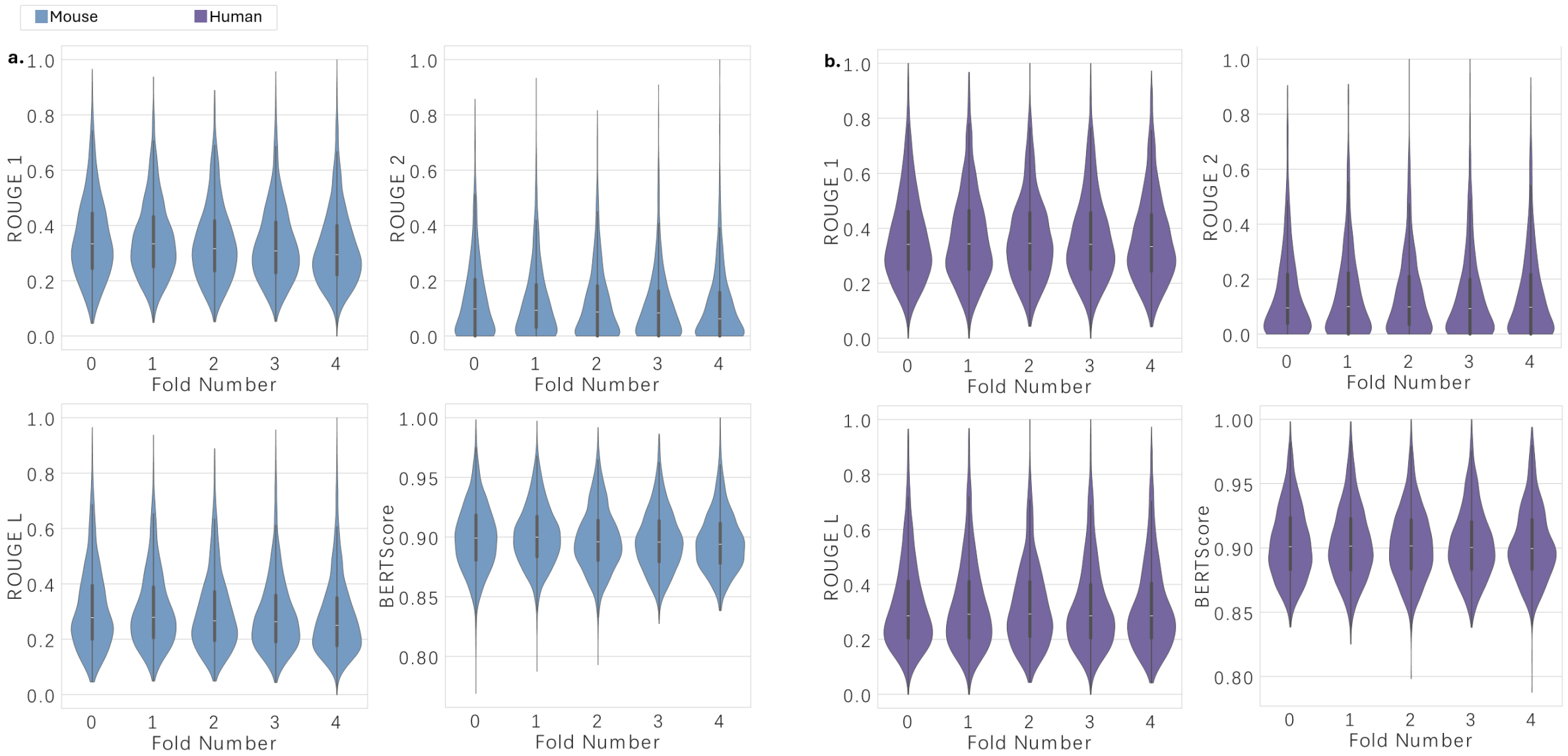

**Supplementary Figs 2. Consistent testing score distributions of the Llama 3 8B model through 5-fold cross-validation for all 4 selected metrics**. To examine the robustness of our method, we separately used mouse (**a**) and human (**b**) data sets from MSigDB[3] to fine-tune the Llama 3 8B model with different partitions of the datasets as the test set during cross-validation. The model's performance remained highly consistent across both species and all 5 folds, indicating good stability. Detailed values and parameters are recorded in **Supplementary Table 4**.

**Supplementary Table 4. Model performance during cross-validation.** Results for **Fig 2c** in the main text and **Supplementary Fig 2**. We recorded the mean values, standard deviations, and 95% confidence intervals for four metrics achieved by GPTON based on Llama 3 8B during 5-fold cross-validation for both mouse and human datasets. The lack of significant differences in mean scores indicates that the model is highly robust and stable.

| Species | Fold | Metric | Mean | Standard deviation | 95% Confidence interval | Sample size (n) |
|---|---|---|---|---|---|---|
| Mouse | 0 | ROUGE 1 | 0.3553 | 0.1526 | [0.3476, 0.3631] | 1484 |
| Mouse | 1 | ROUGE 1 | 0.3517 | 0.1414 | [0.3445, 0.3589] | 1484 |
| Mouse | 2 | ROUGE 1 | 0.3390 | 0.1421 | [0.3318, 0.3463] | 1484 |
| Mouse | 3 | ROUGE 1 | 0.3316 | 0.1388 | [0.3245, 0.3387] | 1484 |
| Mouse | 4 | ROUGE 1 | 0.3217 | 0.1428 | [0.3144, 0.3290] | 1484 |
| Mouse | 0 | ROUGE 2 | 0.1356 | 0.1433 | [0.1283, 0.1429] | 1484 |
| Mouse | 1 | ROUGE 2 | 0.1268 | 0.1340 | [0.1200, 0.1336] | 1484 |
| Mouse | 2 | ROUGE 2 | 0.1206 | 0.1317 | [0.1139, 0.1273] | 1484 |
| Mouse | 3 | ROUGE 2 | 0.1128 | 0.1270 | [0.1064, 0.1193] | 1484 |
| Mouse | 4 | ROUGE 2 | 0.1103 | 0.1304 | [0.1037, 0.1169] | 1484 |
| Mouse | 0 | ROUGE L | 0.3113 | 0.1527 | [0.3035, 0.3191] | 1484 |
| Mouse | 1 | ROUGE L | 0.3095 | 0.1405 | [0.3034, 0.3167] | 1484 |
| Mouse | 2 | ROUGE L | 0.2969 | 0.1408 | [0.2898, 0.3041] | 1484 |
| Mouse | 3 | ROUGE L | 0.2888 | 0.1360 | [0.2818, 0.2957] | 1484 |
| Mouse | 4 | ROUGE L | 0.2805 | 0.1401 | [0.2733, 0.2876] | 1484 |
| Mouse | 0 | BERTScore | 0.9001 | 0.0282 | [0.8986, 0.9015] | 1484 |
| Mouse | 1 | BERTScore | 0.9015 | 0.0259 | [0.9002, 0.9028] | 1484 |
| Mouse | 2 | BERTScore | 0.8983 | 0.0267 | [0.8969, 0.8997] | 1484 |
| Mouse | 3 | BERTScore | 0.8972 | 0.0256 | [0.8959, 0.8985] | 1484 |
| Mouse | 4 | BERTScore | 0.8961 | 0.0252 | [0.8948, 0.8974] | 1484 |
| Human | 0 | ROUGE 1 | 0.3671 | 0.1636 | [0.3587, 0.3755] | 1459 |
| Human | 1 | ROUGE 1 | 0.3686 | 0.1653 | [0.3601, 0.371] | 1459 |
| Human | 2 | ROUGE 1 | 0.3658 | 0.1555 | [0.3578, 0.3738] | 1458 |
| Human | 3 | ROUGE 1 | 0.3665 | 0.1599 | [0.3582, 0.3747] | 1458 |
| Human | 4 | ROUGE 1 | 0.3621 | 0.1620 | [0.3538, 0.3704] | 1458 |
| Human | 0 | ROUGE 2 | 0.1448 | 0.1581 | [0.1367, 0.1529] | 1459 |
| Human | 1 | ROUGE 2 | 0.1476 | 0.1621 | [0.1393, 0.1559] | 1459 |
| Human | 2 | ROUGE 2 | 0.1429 | 0.1512 | [0.1352, 0.1507] | 1458 |
| Human | 3 | ROUGE 2 | 0.1406 | 0.1607 | [0.1324, 0.1489] | 1458 |
| Human | 4 | ROUGE 2 | 0.1457 | 0.1622 | [0.1373, 0.1540] | 1458 |
| Human | 0 | ROUGE L | 0.3250 | 0.1634 | [0.3166, 0.3334] | 1459 |
| Human | 1 | ROUGE L | 0.3250 | 0.1648 | [0.3165, 0.3334] | 1459 |
| Human | 2 | ROUGE L | 0.3228 | 0.1543 | [0.3149, 0.3308] | 1458 |
| Human | 3 | ROUGE L | 0.3221 | 0.1615 | [0.3138, 0.3304] | 1458 |
| Human | 4 | ROUGE L | 0.3207 | 0.1616 | [0.3124, 0.3290] | 1458 |
| Human | 0 | BERTScore | 0.9042 | 0.0285 | [0.9028, 0.9057] | 1459 |
| Human | 1 | BERTScore | 0.9040 | 0.0289 | [0.9025, 0.9055] | 1459 |
| Human | 2 | BERTScore | 0.9035 | 0.0281 | [0.9021, 0.9049] | 1458 |
| Human | 3 | BERTScore | 0.9034 | 0.0281 | [0.9020, 0.9049] | 1458 |
| Human | 4 | BERTScore | 0.9032 | 0.0288 | [0.9017, 0.9047] | 1458 |

**Supplementary Table 5. Model comparison on the Reactome prediction for mouse gene sets (Fig 2f)**. We evaluated the mean performances and 95% confidence intervals across three metrics of GPTON using two different Llama 3 models on the mouse Reactome dataset (N=1243) from MSigDB[3]. These results were compared to the baseline GPT-4, with p-values from one-sided t-tests also reported. The result shows that GPTON exhibits high generalizability across data sets of different origins.

| Model | Species | Metric | Mean | 95% Confidence interval | One-sided t-test p-value |
|---|---|---|---|---|---|
| Llama 3 70B | Mouse | ROUGE 1 | **0.3343** | [0.3270, 0.3416] | 3.8106e-133 |
| Llama 3 70B | Mouse | ROUGE 2 | **0.1063** | [0.0997, 0.1128] | 5.4817e-42 |
| Llama 3 70B | Mouse | ROUGE L | **0.2841** | [0.2768, 0.2914] | 1.8389e-90 |
| Llama 3 8B | Mouse | ROUGE 1 | 0.3128 | [0.3056, 0.3201] | 2.3895e-107 |
| Llama 3 8B | Mouse | ROUGE 2 | 0.08944 | [0.0832, 0.0957] | 2.1814e-26 |
| Llama 3 8B | Mouse | ROUGE L | 0.2635 | [0.2565, 0.2706] | 5.0607e-69 |
| GPT-4 | Mouse | ROUGE 1 | 0.1553 | [0.1440, 0.1666] | N/A |
| GPT-4 | Mouse | ROUGE 2 | 0.0358 | [0.0282, 0.0434] | N/A |
| GPT-4 | Mouse | ROUGE L | 0.1454 | [0.1347, 0.1561] | N/A |

**Supplementary Fig 3. The performance comparison between fine-tuned Llama 3 70B, 8B, and GPT-4 on the human Reactome data set.**

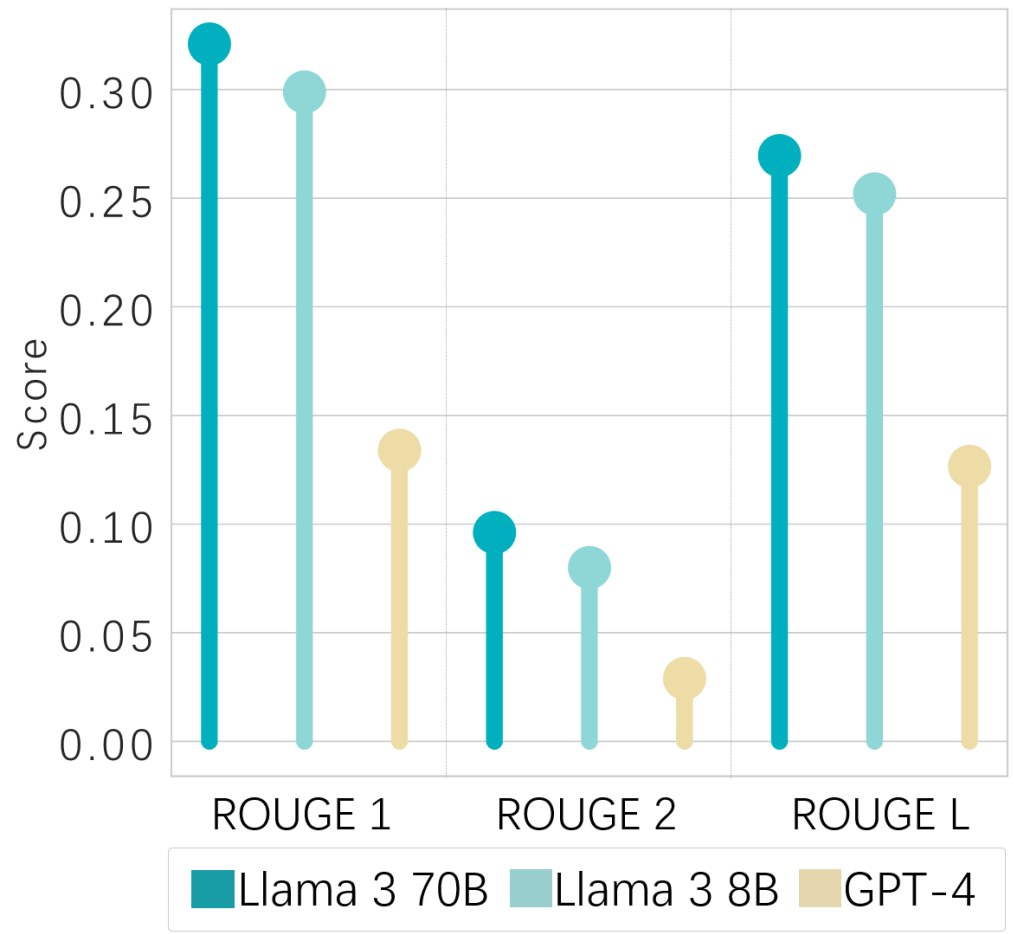

**Supplementary Fig 3. The performance comparison between our fine-tuned models and the baseline GPT-4 model on the human Reactome data set.** Significant improvements over the baseline are observed across all three metrics. The best-performing model, Llama 3 70B, achieves a 19% enhancement in the ROUGE 1 score. Detailed values and parameters are recorded in **Supplementary Table 6**.

**Supplementary Table 6. Model comparison on Reactome prediction on human data set.** Results for **Supplementary Fig 3** that record the superior mean performances and 95% confidence intervals across three metrics of GPTON using two different Llama 3 models on human Reactome data set (N=1670). Their performance is compared to that of the baseline GPT-4, and the p-values of the one-sided t-tests are also recorded.

| Model | Species | Metric | Mean | 95% Confidence interval | One-sided t-test p-value |
|---|---|---|---|---|---|
| Llama 3 70B | Human | ROUGE 1 | **0.3210** | [0.3147, 0.3273] | 2.6391e-218 |
| Llama 3 70B | Human | ROUGE 2 | **0.0961** | [0.0905, 0.1018] | 6.9090e-60 |
| Llama 3 70B | Human | ROUGE L | **0.2696** | [0.2634, 0.2758] | 4.7881e-145 |
| Llama 3 8B | Human | ROUGE 1 | 0.2989 | [0.2929, 0.3049] | 1.0664e-180 |
| Llama 3 8B | Human | ROUGE 2 | 0.0800 | [0.0749, 0.0850] | 1.2602e-39 |
| Llama 3 8B | Human | ROUGE L | 0.2520 | [0.2462, 0.2578] | 1.8318e-199 |
| GPT-4 | Human | ROUGE 1 | 0.1340 | [0.1252, 0.1427] | N/A |
| GPT-4 | Human | ROUGE 2 | 0.0289 | [0.0234, 0.0345] | N/A |
| GPT-4 | Human | ROUGE L | 0.1266 | [0.1182, 0.1350] | N/A |

**Supplementary Fig 4. The line plots of the performance of Llama 3 8B with reduced training sample sizes.**

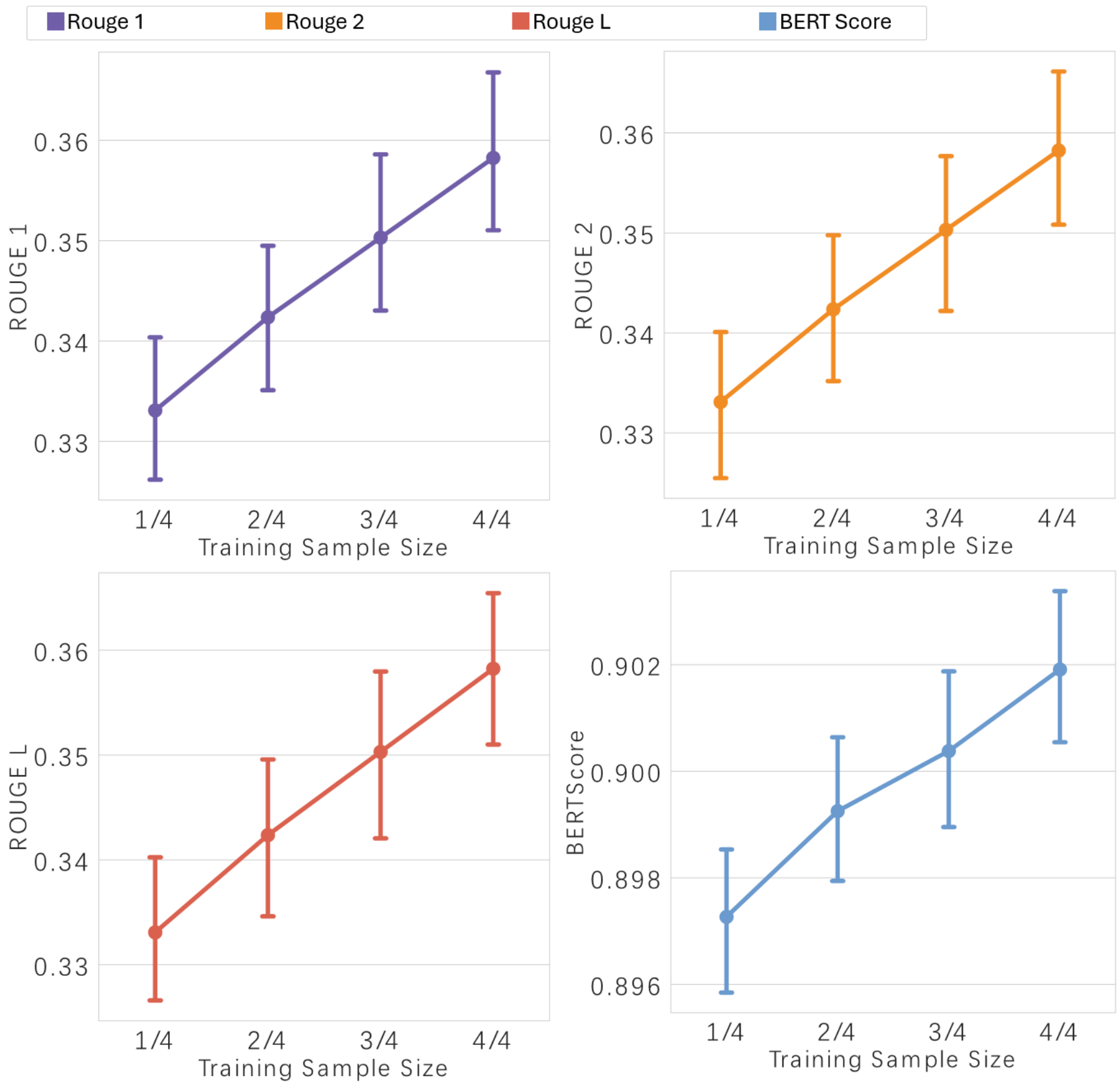

**Supplementary Fig 4. The performance of the Llama 3 8B model improves along with the increased sample size used for fine-tuning**. We sampled different fractions of the mouse gene training set to fine-tune the model separately to evaluate the effect of training sample size on model performance. The figures show that performance improves across all metrics as sample size increases, informing potential future research. The error bars represent 95% confidence intervals. Detailed values and parameters are recorded in **Supplementary Table 7.**

**Supplementary Table 7. Model performance observation with different training sample sizes using Llama 3 8B.** Detailed results for **Fig 2g** in the main text and **Supplementary Fig 4**. The results show increasing mean performance of GPTON, using the fine-tuned Llama 3 8B model, across all four metrics as training sample size increases. The test sample set is the verbalized mouse data set (total N=1489). This indicates that more training samples could further enhance the model's performance, providing insights for future research.

| Training sample fraction | Metric | Mean | 95% Confidence interval |
|---|---|---|---|
| 1/4 | ROUGE 1 | 0.3331 | [0.3260, 0.3402] |
| 1/2 | ROUGE 1 | 0.3424 | [0.3349, 0.3499] |
| 3/4 | ROUGE 1 | 0.3503 | [0.3425, 0.3581] |
| 4/4 | ROUGE 1 | **0.3582** | [0.3505, 0.3660] |
| 1/4 | ROUGE 2 | 0.1128 | [0.1063, 0.1194] |
| 1/2 | ROUGE 2 | 0.1257 | [0.1188, 0.1326] |
| 3/4 | ROUGE 2 | 0.1333 | [0.1259, 0.1408] |
| 4/4 | ROUGE 2 | **0.1366** | [0.1291, 0.1441] |
| 1/4 | ROUGE L | 0.2903 | [0.2833, 0.2974] |
| 1/2 | ROUGE L | 0.3011 | [0.2937, 0.3085] |
| 3/4 | ROUGE L | 0.3093 | [0.3016, 0.3171] |
| 4/4 | ROUGE L | **0.3142** | [0.3066, 0.3219] |
| 1/4 | BERTScore | 0.8973 | [0.8959, 0.8986] |
| 1/2 | BERTScore | 0.8993 | [0.8979, 0.9006] |
| 3/4 | BERTScore | 0.9004 | [0.8989, 0.9018] |
| 4/4 | BERTScore | **0.9019** | [0.9005, 0.9033] |